\documentclass[preprint,pre,eqsecnum,showpacs]{revtex4}
\usepackage[centertags]{amsmath}
\usepackage{amssymb}
\usepackage{amsthm}
\usepackage{graphicx}
\usepackage{epsfig}
\usepackage{mathrsfs}
\usepackage[T1]{fontenc}
\begin{document}

\title{Coherent state of a weakly interacting ultracold Fermi gas}
\author{Arnab Ghosh, Sudarson Sekhar Sinha and Deb Shankar Ray
{\footnote{Email Address: pcdsr@iacs.res.in}}} \affiliation{Indian
Association for the Cultivation of Science, Jadavpur, Kolkata 700
032, India.}

\begin{abstract}
\indent We examine the weakly interacting atoms in an ultracold
Fermi gas leading to a state of macroscopic coherence, from a theoretical
perspective. It has been shown that this state can be
described as a fermionic coherent state. These coherent states are the
eigenstates of fermionic annihilation operators, the eigenvalues
being anti-commuting numbers or Grassmann numbers. By exploiting the
simple rules of Grassmann algebra and a close kinship between
relations evaluated for more familiar bosonic fields and those for
fermionic fields, we derive the thermodynamic limit, the spontaneous
symmetry breaking and the quasi-particle spectrum of the fermionic
system.
\end{abstract}
\maketitle
\newpage
\section{{INTRODUCTION}}
\indent Dilute, ultracold, trapped atomic gases have served as model
systems for exploring quantum phenomena at the fundamental level
\cite{1,2,3,4}. One of the basic issues is the condensation of a
Fermi gas. Important experimental advances on alkali atoms
$^{6}\text{Li}$ and $^{40}\text{K}$ \cite{5,6} have been made in
this regard to study the interaction between the fermions. Since the
interaction between cold atoms characterized by s-wave scattering
length can be controlled by magnetic field Feshbach resonances, it
has been possible to realize both the repulsive (BEC) and the
attractive (BCS) regimes and their crossover. In spite of a very
large number of theoretical papers published over a decade, a
complete analytic solution of the many-body problem along the
BEC-BCS crossover still remains illusive \cite{7,8,9}. However, the
use of mean field theory, pseudopotential and consideration of
fluctuations around mean field has led interesting advancement in
understanding the basic physics of ultracold atomic Fermi gases. For
details we refer to the topical review by Giorgini, Pitaevskii
and Stringari \cite{9}.\\
\indent To capture the essential physics around the crossover it is
necessary to comprehend the formation of Cooper pairs or molecule or
BEC condensate or condensation of fermionic atom pairs
\cite{10,11,12,13,14}. A close look into this aspect reveals that
the description of ``Fermi condensate'' around this crossover is still
somewhat incomplete \cite{14}. This is because of the fact that a simple
two-body physics of the resonance on the attractive side of
interaction does not support a weakly bound molecular state -- a
point emphasized by Jin $\textit{et. al.}$ \cite{12}. This implies
that many-body effects must prevail in characterizing such a state. 
In the present study we focus on the description of 
this state in a weakly interacting Fermi gas. Pauli principle
rules out the possibility of macroscopic occupation of a single quantum
energy state of fermions. A question, however, remains, whether it is
possible to realize a state of macroscopic coherence among them. This
is based on exploitation, following Cahill and Glauber \cite{15}, of
the close parallels between the families of the quasi-probability
phase space density functions for bosonic and fermionic fields. A
key element of the formulation is the fermionic coherent state
defined as a displaced state where the displacement operator acts on
the vacuum state. The transformation with the displacement operator
displaces a fermionic field operator over an anti-commuting number
or Grassmann number \cite{15,16,17,18}. Fermionic coherent state is
an eigenstate of annihilation operator, eigenvalues being the
Grassmann numbers. This is reminiscent of the harmonic oscillator
coherent state for which the displacement operator displaces the
bosonic field operator over a classical commuting number
\cite{19,20,21}. Our method for description of the macroscopic state
of weakly interacting fermions is based on the fermionic coherent
state and the associated algebra of the anti-commuting numbers,
which have no clasical analog \cite{15,16,17,18}. They form the basis for
demonstration of the thermodynamic limit, spontaneous symmetry
breaking and the quasi-particle spectrum of the atoms comprising the
coherent state. A major goal of the approach is to show that
it is possible to achieve a coherent state of weakly interacting fermions
for which the scattering length can be either positive or negative.
And the state can exist as independent entity irrespective of the
details of Cooper pairing or molecule formation.\\
\indent The paper is organized as follows: In Sec.II we briefly
introduce the basic model for weakly interacting many body systems
under ultracold condition. Since
Grassmann numbers play a crucial role in the formulation of the
problem we briefly review in Sec.III the relevant aspects of
Grassmann algebra before introducing the fermionic coherent state
following Cahill and Glauber \cite{15}. That the state of macroscopic
coherence can be realized as a fermionic coherent state is the main theme of this
section. We further examine the aspect of thermodynamic limit and
spontaneous symmetry breaking . In Sec.IV we derive the quasi-particle spectrum
using Bogoliubov transformation. The paper is concluded in
Sec.V.
\section{{Weakly Interacting Fermi gas}}
\indent To begin with we consider a dilute gas of fermionic atoms.
The interaction between them is as usual nonzero only when the range
of inter-particle interaction $r_{0}$ is much smaller than the
average inter-particle distance $d$, i.e.,
$r_{0}<<d=\rho^{-1/3}=(\frac{N}{V})^{1/3}$. This also ensures that
the properties of the system can be expressed in terms of a single
parameter, scattering length, $a$. The diluteness criterion in terms
s-wave scattering length can be expressed as $|a|\rho^{1/3}<<1$. We
also assume that the temperature of the dilute gas is so low that
the momentum $q$ as distributed thermally is much smaller than the
characteristic momentum $q_{c}=\hbar/r_{0}$. The scattering
amplitude is independent of momentum $q$ when $q<<q_{c}$. While the
temperature for onset of quantum degeneracy goes as $\rho^{2/3}$,
temperature for BEC-BCS crossover is of the order of Fermi
temperature for a range of dimensionless $k_{\textbf{F}}|a|$ values,
where $k_{\textbf{F}}$ corresponds to Fermi wave vector
\cite{7,8,9}.\\
\indent The Hamiltonian of the system can be expressed in terms of
the field operator $\hat{\psi}(r)$ as
\begin{eqnarray} \label{3.2}
\hat{\textbf{H}}=\int\left(\frac{\hbar^{2}}{2m}
\nabla\hat{\psi}^{\dag}(r)\nabla\hat{\psi}(r)\right)dr
+\frac{1}{2}\int\hat{\psi}^{\dag}(r)\hat{\psi}^{\dag}(r')V(r-r')\hat{\psi}(r)\hat{\psi}(r')dr\;dr'
\end{eqnarray}
where $V(r-r')$ is the two-body potential. For the moment we have
not included the external fields. In absence of external fields, the
$N$ particles move only as a result of their mutual interactions.
The field operator $\hat{\psi}(r)$ annihilates a particle at a
position $r$ and can be expressed as
$\hat{\psi}(r)=\sum_{i}\phi_{i}(r)\hat{a}_{i}$.
$\hat{a}_{i}(\hat{a}^{\dag}_{i})$ is the annihilation (creation)
operator of a particle in the single particle state $\phi_{i}(r)$
and these operators obey anti-commutation relations
\begin{eqnarray} \label{3.2}
\{\hat{a}_{i},\hat{a}^{\dag}_{j}\}=\delta_{ij};\;\;\;
\{\hat{a}_{i},\hat{a}_{j}\}=\{\hat{a}^{\dag}_{i},\hat{a}^{\dag}_{j}\}=0
\end{eqnarray}
The wave functions $\phi_{i}(r)$ satisfy orthonormal condition
\begin{eqnarray} \label{3.2}
\int\phi^{*}_{i}(r)\phi_{j}(r)=\delta_{ij}
\end{eqnarray}
The field operator then follows anti-commutation relation
$\{\hat{\psi}(r),\hat{\psi}^{\dag}(r')\}=\sum_{i}\phi_{i}(r)\phi^{*}_{i}(r')\\=\delta(r-r')$.
Now expressing the field operator $\hat{\psi}(r)$ and the
interaction term, $V(r,r')\;(\equiv V(r-r'))$ as
$\hat{\psi}(r)=\sum_{k}\hat{a}_{k}\frac{e^{ik.r}}{\sqrt{V}}$ and
$V_{q}=\int V(r)e^{-\frac{i.q.r}{\hbar}}dr$, we may simplify the
Hamiltonian as
\begin{eqnarray} \label{3.2}
\hat{\textbf{H}}=\sum_{k}\frac{\hbar^{2}k^{2}}{2m}\hat{a}^{\dag}_{k}\hat{a}_{k}+\frac{1}{2V}
\sum_{k_{1},k_{2},q}V_{q}\hat{a}^{\dag}_{k_{1}+q}\hat{a}^{\dag}_{k_{2}-q}\hat{a}_{k_{2}}\hat{a}_{k_{1}}
\end{eqnarray}
\indent We take into account of the momentum conservation during a
two-fermion interaction such that every single term within the
summation describes the annihilation of a pair of fermionic atoms
and the creation of another pair. Since the description of the
macroscopic properties of the gas does not require the detailed form
of the two-body interaction, microscopic potential $V(r)$ may be
replaced by a soft, effective potential $V_{eff}(q)$ which, in turn,
gives the low momentum value of its Fourier transform
$V_{0}$. For such small momenta $q<<q_{c}=\hbar/r_{0}$ we are allowed to
consider $V_{0}$
or the s-wave scattering length $`a$' becomes independent of its momentum $q$ \cite{7,8,9}.\\
\indent The Hamiltonian therefore reduces to
\begin{eqnarray} \label{3.2}
\hat{\textbf{H}}=\sum_{k}\frac{\hbar^{2}k^{2}}{2m}\hat{a}^{\dag}_{k}\hat{a}_{k}+\frac{V_{0}}{2V}
\sum_{k_{1},k_{2},q}\hat{a}^{\dag}_{k_{1}+q}\hat{a}^{\dag}_{k_{2}-q}\hat{a}_{k_{2}}\hat{a}_{k_{1}}
\end{eqnarray}
This is the starting Hamiltonian for weakly interacting Fermi atoms
and henceforth will be used in the following sections. Finally we
emphasize that the Hamiltonian considered here (Eqs. 2.1-2.5)
follows Greiner \cite{21} and does not refer to specific spin
states. This is an important point of departure from many
theoretical approaches to quantum Fermi liquid \cite{7} and BCS-BEC crossover \cite{7,8,9}.\\
\indent To proceed further for a systematic description we first separate out the
field operator $\hat{\psi}(r)$ into the coherent term
($i=\textbf{F}$) and the incoherent term ($i\neq \textbf{F}$) as;
\begin{eqnarray} \label{3.2}
\hat{\psi}(r)=\phi_{\textbf{F}}(r)\hat{a}_{\textbf{F}}+\sum_{i\neq
\textbf{F}}\phi_{i}(r)\hat{a}_{i}
\end{eqnarray}
\indent We do this separation by taking a hint from Bogoliubov
approximation \cite{22} for bosons and the close parallelism between
the expressions developed for bosonic and fermionic fields by Cahill
and Glauber \cite{15}. We may note that an implementation of
Bogoliubov approximation \cite{22} for fermionic fields requires
replacement of $\hat{a}_{\textbf{F}}$ and
$\hat{a}^{\dag}_{\textbf{F}}$ by anti-commuting numbers, say
$y_{\textbf{F}}$ and $y^{*}_{\textbf{F}}$ \cite{15,16,17,18}. One can always multiply these numbers by a numerical phase factor as $y_{\textbf{F}}e^{i\theta}$ and $y^{*}_{\textbf{F}}e^{-i\theta}$ without changing any physical property. This phase $`\theta$' plays, as we will see, a major role in characterizing the coherence in the weakly interacting ultracold Fermi gas. This reflects the gauge symmetry exhibited by all the physical equations of the problem. Making an explicit choice for the value of the phase actually corresponds to a formal breaking of gauge symmetry.\\
\indent Equivalently the Bogoliubov approximation for fermions is equivalent
to treating the macroscopic component
$\phi_{\textbf{F}}(r)\hat{a}_{\textbf{F}}$ as a ``classical''
non-commuting field $\psi_{\textbf{F}}(r)$ so that $\hat{\psi}(r)$
may be rewritten as
\begin{eqnarray} \label{3.2}
\hat{\psi}(r)=\psi_{\textbf{F}}(r)+\delta\hat{\psi}(r)
\end{eqnarray}
\indent The above ansatz for the fermionic field operator can be
interpreted as the expectation value $\langle\hat{\psi}(r)\rangle$
different from zero. This is not possible if the coherent state is
a particle number eigenstate. Although such
coherent states are well-known for bosonic fields and have been the
basis for understanding BEC \cite{23,24,25,26,27}, an extension of
the scheme to their fermionic counterpart is not straight-forward.
The primary reason may be traced to a basic issue. Since fermions
anti-commute their eigenvalues must be anti-commuting numbers, as
pointed out by Schwinger \cite{16}. Such numbers are Grassmann
numbers which can be dealt with by simple techniques of Grassmann
algebra \cite{17,18}. In what follows in the next section we make a
little digression on this algebra centering around fermionic
coherent state following Cahill and Glauber \cite{15} which forms the integral 
part of the description of the interacting fermions near the crossover.\\
\indent Although an important guideline of the present formulation
is the close parallels between some aspects of bosons and fermions
it is important to highlight the other essentials. The following digression
may be interesting.  While for a bosonic gas the condensed phase is formed
by the particles in the zero energy ground state, in the fermionic gas the particles may
``condense'' \cite{28} in the Fermi level. Such a prediction was made many decades
ago by Kothari and Nath \cite{28} in course of examination of Born's
reciprocity principle \cite{29}. In the present scenario, we note
that the BEC-BCS crossover temperature for a Fermi gas is less than
or equal to the Fermi temperature \cite{10,11,12,13,14}, a point
that hints towards this assertion. In what follows we will show that a 
macroscopic state of weakly interacting
ultracold fermions can give rise to a fermionic coherent state.
\section{{Grassmann algebra and Fermionic coherent states; A connection to 
Macroscopic coherence of Fermions:}}
\subsection{{Grassmann variables and their properties:}}
\indent We now summarize some of the properties of the
anti-commuting classical variables relevant for our future
discussions. These variables can be treated within the scope of
Grassmann algebra, which are well studied in mathematics and field
theory \cite{17}. They possess very uncommon properties.\\
\indent Let, $\textbf{y}=\{y_{i}\}, i=1,2,.....n.$ define a set of
generators which satisfy anti-commutation properties,
\begin{eqnarray} \label{3.1}
y_{i}y_{j}+y_{j}y_{i}\equiv\{y_{i},y_{j}\}=0
\end{eqnarray}
This, in particular, implies that $y_{i}^{2}=0$, for any given
$`i$'. Since the square of every Grassmann monomial vanishes, a nonzero
Grassmann monomial can not be an ordinary real, imaginary, or
complex number. In other words, Grassmann variables are nilpotent, an
important property for the treatment of fermions \cite{18}. The
anti-commuting numbers $y_{i}$ and their complex conjugates
$y^{*}_{i}$ are independent numbers and satisfy
\begin{eqnarray} \label{3.3-3.6}
\{y_{i}, y^{*}_{i}\}&=&0
\end{eqnarray}
They also anti-commute with their fermionic operators
\begin{eqnarray} \label{3.7-3.8}
\{y_{i}, \hat{a}_{j}\}=0\;;\;\;\;\;\;\;\;\;\;\;\{y_{i},
\hat{a}^{\dag}_{j}\}=0
\end{eqnarray}
And hermitian conjugation reverses the order of all fermionic
quantities, both the operators and the anti-commuting numbers. For
instance, we have
\begin{eqnarray} \label{3.9}
(\hat{a}_{1} y_{2} \hat{a}^{\dag}_{3}y^{*}_{4})^{\dag}=y_{4}
\hat{a}_{3} y^{*}_{2} \hat{a}^{\dag}_{1}
\end{eqnarray}
\indent An analytic function of only one Grassmann variable can be
expressed as a simple Taylor expansion
\begin{eqnarray} \label{3.10}
f(y)=f_{0}+f_{1}y\;\;\;\;\;\;(\text{since $y^{2}=0$})
\end{eqnarray}
Thus exponential function for Grassmann variables has only two
terms
\begin{eqnarray} \label{3.25}
e^{y}=\sum^{\infty}_{k=0}\frac{y^{k}}{k!}=1+y\;\;\;\;(\text{since
$y^{2}=0$})
\end{eqnarray}
Therefore for a function of single Grassmann variable, integration
(integration due to Berezin \cite{17}) is identical to
differentiation.
The fundamental rules of integration over the complex
Grassmann variables are as follows;
\begin{eqnarray} \label{3.18-3.21}
\int dy_{i}=0\;\;\;\;\;\;\;
\int dy^{*}_{i}=0\\
\int dy_{i}\;y_{j}=\delta_{ij}\;\;\;\;\; \int
dy^{*}_{i}\;y^{*}_{j}=\delta_{ij}
\end{eqnarray}
This difference between ordinary variables and
Grassmann variables has many interesting consequences. We are
typically concerned with pairs of anti-commuting variables $y_{i}$
and $y^{*}_{i}$, and for such pairs we will confine ourselves to the
notation
\begin{eqnarray} \label{3.23}
\int d^{2}y_{i} =\int dy^{*}_{i}dy_{i}
\end{eqnarray}
in which the differential of the conjugated variable $dy^{*}_{i}$
comes first and we keep in mind that
\begin{eqnarray} \label{3.24}
dy^{*}_{i}dy_{i}=-dy_{i}dy^{*}_{i}
\end{eqnarray}
We can write the multiple integrals over such sets as
\begin{eqnarray} \label{3.24}
\int d^{2}\textbf{y}\equiv \int \prod_{i}dy^{*}_{i}dy_{i}
\end{eqnarray}
\subsection{{Fermionic coherent states:}}
\indent We are now in a position to introduce the fermionic coherent
states. In analogy to harmonic oscillator coherent state
$|\boldsymbol\alpha\rangle$ \cite{19} defined as a displaced state
where the displacement operator
$\hat{\texttt{D}}(\boldsymbol\alpha)=
\exp\left(\sum_{i}(\alpha_{i}\hat{a}^{\dag}_{i}-\alpha^{*}_{i}\hat{a}_{i})\right)$
acts on the vacuum $|0\rangle$ as
$|\boldsymbol\alpha\rangle=\hat{\texttt{D}}(\boldsymbol\alpha)|0\rangle$,
$\{\alpha_{i}\}$ being a set of complex numbers \cite{19,20,21}, it
is possible to construct a displacement operator for fermions \cite{15} as
\begin{eqnarray} \label{3.33}
\hat{{\textbf{D}}}(\textbf{y})=\exp\left(\sum_{i}(\hat{a}^{\dag}_{i}y_{i}-y^{*}_{i}\hat{a}_{i})\right)
\end{eqnarray}
for a set of $\textbf{y}=\{y_{i}\}$ Grassmann variables and the
fermionic coherent state can then be constructed by the action of
this displacement operator (Eq. 3.12) on the vacuum state as
\begin{eqnarray} \label{3.33}
|\textbf{y}\rangle=\hat{{\textbf{D}}}(\textbf{y})|0\rangle
\end{eqnarray}
\subsubsection{{Displacement operator and its properties:}}
\indent An important property of the Grassmann variables is that on
multiplication with fermionic annihilation or creation operators,
their anti-commutivity cancels that of the operators. Thus the
operators $\hat{a}^{\dag}_{i}y_{i}$ and $y^{*}_{j}\hat{a}_{j}$
simply commute for $i\neq j$ \cite{15}. Therefore the displacement operator
may be written as the product
\begin{eqnarray} \label{3.33}
\hat{{\textbf{D}}}(\textbf{y})&=&\prod_{i}\exp\left(\hat{a}^{\dag}_{i}y_{i}-y^{*}_{i}\hat{a}_{i}\right)\nonumber\\
&=&\prod_{i}\left[1+\hat{a}^{\dag}_{i}y_{i}-y^{*}_{i}\hat{a}_{i}
+\left(\hat{a}^{\dag}_{i}\hat{a}_{i}-\frac{1}{2}\right)y^{*}_{i}y_{i}\right]
\end{eqnarray}
\indent Similarly the annihilation operators $\hat{a}_{k}$ commutes
with all the operators $\hat{a}^{\dag}_{i}y_{i}$ and
$y^{*}_{i}\hat{a}_{i}$ for $k\neq i$, and so we may compute the
displaced annihilation operator by ignoring all modes except the
k-th one
\begin{eqnarray} \label{3.33}
\hat{{\textbf{D}}}^{\dag}(\textbf{y})\hat{a}_{k}\hat{{\textbf{D}}}(\textbf{y})
&=&\prod_{i}\exp\left(y^{*}_{i}\hat{a}_{i}-\hat{a}^{\dag}_{i}y_{i}\right).\hat{a}_{k}.\prod_{j}
\exp\left(\hat{a}^{\dag}_{j}y_{j}-y^{*}_{j}\hat{a}_{j}\right)\nonumber\\
&=&\exp\left(y^{*}_{k}\hat{a}_{k}-\hat{a}^{\dag}_{k}y_{k}\right).\hat{a}_{k}.
\exp\left(\hat{a}^{\dag}_{k}y_{k}-y^{*}_{k}\hat{a}_{k}\right)\nonumber\\
&=&\hat{a}_{k}+y_{k}
\end{eqnarray}
Similarly we have
\begin{eqnarray} \label{3.33}
\hat{{\textbf{D}}}^{\dag}(\textbf{y})\hat{a}^{\dag}_{k}\hat{{\textbf{D}}}(\textbf{y})
&=&\hat{a}^{\dag}_{k}+y^{*}_{k}
\end{eqnarray}
\subsubsection{{Properties of the coherent states:}}
\indent By using the displacement relation [Eq (3.15)] we may show
that the coherent state is an eigenstate of every annihilation
operator $\hat{a}_{k}$;
\begin{eqnarray} \label{3.33}
\hat{a}_{k}|\textbf{y}\rangle=\hat{a}_{k}\hat{{\textbf{D}}}(\textbf{y})|0\rangle
&=&\hat{{\textbf{D}}}(\textbf{y})\hat{{\textbf{D}}}^{\dag}(\textbf{y})
\hat{a}_{k}\hat{{\textbf{D}}}(\textbf{y})|0\rangle\nonumber\\
&=&\hat{{\textbf{D}}}(\textbf{y})(\hat{a}_{k}+y_{k})|0\rangle\nonumber\\
&=&y_{k}\hat{{\textbf{D}}}(\textbf{y})|0\rangle=y_{k}|\textbf{y}\rangle
\end{eqnarray}
The adjoint of the coherent state $|\textbf{y}\rangle$ can be
similarly defined as
$\langle\textbf{y}|\hat{a}^{\dag}_{k}=\langle\textbf{y}|y^{*}_{k}$.\\
The inner product of two coherent states is
\begin{eqnarray} \label{3,38}
\langle
\textbf{y}'|\textbf{y}\rangle&=&\exp\left(\sum_{i}y'^{*}_{i}y_{i}
-\frac{1}{2}(y'^{*}_{i}y'_{i}+y^{*}_{i}y_{i})\right)
\end{eqnarray}
and using the completeness properties of the coherent states, any
arbitrary coherent state $|\textbf{y}'\rangle$ can be expanded \cite{15,18} as
\begin{eqnarray} \label{3.39}
|\textbf{y}'\rangle=\int d^{2}\textbf{y}\langle
\textbf{y}|\textbf{y}'\rangle|\textbf{y}\rangle
\end{eqnarray}
which immediately follows from the resolution of identity
\begin{eqnarray} \label{3.40}
\int d^{2}\textbf{y}|\textbf{y}\rangle\langle \textbf{y}|=\textbf{I}
\end{eqnarray}
\subsection{{Weakly interacting Fermi gas in coherent state representation:}}
\indent The key idea behind the present formulation 
is whether a state of macroscopic coherence of an weakly interacting 
Fermi gas can be achieved as a fermionic coherent state. Such an 
approach gives the result which is
equivalent to that of the problem of fixed number of particles $N$
in the limit $N\longrightarrow\infty$ for the BEC case
\cite{23,24,25,26,27}. In the same spirit, we can extend the
coherent state approach to its fermionic counterpart. Before
proceeding further, we discuss the physical states and operators,
which will be helpful for the future development.
\subsubsection{{Physical states and operators :}}
\indent According to Cahill and Glauber, \cite{15} a state $|\psi\rangle$ is
physical if it changes at most by a phase when subjected to a
rotation of angle $2\pi$ about any axis,
\begin{eqnarray} \label{3.40}
\textbf{U}(\hat{n},2\pi)|\psi\rangle=e^{i\theta}|\psi\rangle
\end{eqnarray}
\indent Since, fermions carry half-integer spin, a state of any odd
number of fermions changes by the phase factor -1, while even number
of fermions are invariant under such $2\pi$ rotation. Thus physical
states are linear combinations of states with even number of
fermions or linear combination of odd number of fermions. On the
other hand, a linear combination of states $|0\rangle$ and
$|1\rangle$ like $\frac{1}{\sqrt{2}}(|0\rangle+|1\rangle)$ must be
unphysical since it changes to a different state under $2\pi$ rotation.\\
\indent An operator is physical if it maps physical states onto
physical states. Physical operators are either even or odd. In all
physical contexts, if we consider
$\hat{N}=\sum_{k}\hat{a}^{\dag}_{k}\hat{a}_{k}$ as the total fermion
number which is conserved, then any state arising from the
eigenstate of $\hat{N}$ must remain an eigenstate of $\hat{N}$. This
can be derived on the basis of $\textbf{U}(1)$ invariance of all the
interactions under the transformation
$\hat{U}(\theta)=e^{i\theta\hat{N}}$. We have the transformation
on $\hat{a}_{i}$ and $\hat{a}^{\dag}_{i}$ as follows;
\begin{eqnarray} \label{3.39}
e^{-i\theta\hat{N}}\hat{a}_{i}e^{i\theta\hat{N}}=e^{i\theta}\hat{a}_{i}
\end{eqnarray}
and
\begin{eqnarray} \label{3.39}
e^{-i\theta\hat{N}}\hat{a}^{\dag}_{i}e^{i\theta\hat{N}}=e^{-i\theta}\hat{a}^{\dag}_{i}
\end{eqnarray}
for fermion conserving interactions that involve $\hat{a}_{k}$ and
$\hat{a}^{\dag}_{k}$; and the phase factors get cancel out.
We now emphasize that the coherent states undergo a simple
change under this transformation
\begin{eqnarray} \label{3.39}
\hat{U}(\theta)|\textbf{y}\rangle=e^{i\theta\hat{N}}|\textbf{y}\rangle=
|e^{i\theta}\textbf{y}\rangle,
\end{eqnarray}
However the scalar product remains invariant
\begin{eqnarray} \label{3.39}
\langle
e^{i\theta}\textbf{y}|e^{i\theta}\textbf{y}\rangle=\langle\textbf{y}|\textbf{y}\rangle
\end{eqnarray}
\subsubsection{{Spontaneous symmetry breaking:}}
\indent We now note that one can always multiply the coherent state
by an arbitrary phase factor $e^{i\theta}$ without changing any
physical property (as far as physical states are concerned). This is
the manifestation of gauge symmetry in the problem. Physically, the
lack of a force responsible for phase stabilization of the system is
the origin for the random phase of a condensate. However, when one
refers to Fermi systems the macroscopic system is expected to
choose spontaneously a particular phase $``\theta$''. Making an
explicit choice for the phase $``\theta$'' in spite of the lack of a
preferred phase value (referred to as a spontaneous breaking of
gauge symmetry) implies that the macroscopic state is in or close to a
coherent state. From the symmetry point of view, the situation is
quite interesting and can be further
elaborated as follows.\\
\indent The states $|\textbf{y}\rangle$ are not invariant under the
number operator $\hat{N}=\sum_{k}\hat{a}^{\dag}_{k}\hat{a}_{k}$,
while the Hamiltonian (Eq. 2.5) commutes with $\hat{N}$, i.e.,
\begin{eqnarray} \label{3.39}
e^{i\theta\hat{N}}|\textbf{y}\rangle=|e^{i\theta}\textbf{y}\rangle;\;\;\;\;\;
e^{i\theta\hat{N}}\hat{\textbf{H}}e^{-i\theta\hat{N}}=\hat{\textbf{H}}
\end{eqnarray}
The operator $e^{i\theta\hat{N}}$ applied to $|\textbf{y}\rangle$
produces a state with the same energy, with a phase shifted by
$``\theta$''. Since the overlap of coherent states obey $\langle
\textbf{y}'|\textbf{y}\rangle=\exp\left(\sum_{i}y'^{*}_{i}y_{i}
-\frac{1}{2}(y'^{*}_{i}y'_{i}+y^{*}_{i}y_{i})\right)$ [Eq. (3.18)],
any two different states $|\textbf{y}\rangle$, $|\textbf{y}'\rangle$
with different phase factors, $|\textbf{y}\rangle$,
$|e^{i\theta}\textbf{y}\rangle$, parameterized by a phase variable
$0<\theta<2\pi$, are macroscopically distinct. While the microscopic
Hamiltonian $(\hat{\textbf{H}})$ [Eq. 2.5] has global
$\textbf{U}(1)$ symmetry, the state does not possess such symmetry,
since adding a phase factor to the state $|\textbf{y}\rangle$
produces a different state altogether. 
\subsubsection{{Population fluctuation and phase stabilization:}}
\indent One can easily calculate the population fluctuation in the
coherent state given by Eq. (3.13) as
\begin{eqnarray} \label{3.39}
\langle\Delta\hat{N}^{2}\rangle=
\langle\hat{N}^{2}\rangle-\langle\hat{N}\rangle^{2}=\sum_{k}y^{*}_{k}y_{k}
=\langle\hat{N}\rangle
\end{eqnarray}
Here we have to use the property \cite{15} of the fermionic number
operator $\hat{N}=\sum_{k}\hat{a}^{\dag}_{k}\hat{a}_{k}$ and the nilpotency of the Grassmann variables
i.e., $(y^{*}_{k}y_{k})^{2}=0$, for each k.\\
\indent Then the phase fluctuation in the coherent state is given by
\begin{eqnarray} \label{3.39}
\langle\Delta\hat{\theta}^{2}\rangle=\frac{1}{4\langle\Delta\hat{N}^{2}\rangle}
=\frac{1}{4\langle\hat{N}\rangle}=\frac{1}{4\sum_{k}y^{*}_{k}y_{k}}
\end{eqnarray}
Since $\hat{N}$ and $\hat{\theta}$ are the canonically conjugate
pairs, the phase is stabilized
$\langle\Delta\hat{\theta}^{2}\rangle<<1$, only at the cost of the
enhanced fluctuation in population i.e.,
$\langle\Delta\hat{N}^{2}\rangle>>1$. Since the coherent state can
be expanded as a coherent superposition of the particle number
eigenstates [Eq. (3.13)], the constructive and destructive
interferences among different number eigenstates result in the
stabilized phase but with finite particle number noise. Population
fluctuation and phase stabilization are the typical signatures of
spontaneous symmetry breaking \cite{27}.
\subsubsection{{Thermodynamic limit; Grassmann-Bogoliubov approximation:}}
\indent Once a macroscopic state of interacting fermionic atoms is
realized as the coherent state, neither $\hat{a}_{\textbf{F}}$ nor
$\hat{a}^{\dag}_{\textbf{F}}$ annihilates the state. Since, we are
interested in the behaviour of a gas of fermionic atoms, i.e., in
the large particle number and volume, it is necessary to consider
the so-called thermodynamic limit $N\longrightarrow\infty$,
$V\longrightarrow\infty$ but with constant density
$\rho=\frac{N}{V}$. In this limit, the anti-commutation relation
between fermionic operators $\hat{a}_{\textbf{F}}$ and
$\hat{a}^{\dag}_{\textbf{F}}$ becomes
\begin{eqnarray} \label{3.39}
\frac{\{\hat{a}_{\textbf{F}},\hat{a}^{\dag}_{\textbf{F}}\}}{V}
=\frac{\hat{a}_{\textbf{F}}\hat{a}^{\dag}_{\textbf{F}}
+\hat{a}^{\dag}_{\textbf{F}}\hat{a}_{\textbf{F}}}{V}=\frac{1}{V}\longrightarrow0
\;\;\;\;\;\;\;\;\;(\text{when}\;V\longrightarrow\infty)
\end{eqnarray}
\indent In the limit, $V\longrightarrow\infty$, we are allowed to
forget the operator character of $\hat{a}_{\textbf{F}}$ and
$\hat{a}^{\dag}_{\textbf{F}}$ and they can be replaced by numbers,
i.e., we obtain the ``classical'' limit of the fermionic operators.
To make this point explicit, let us define
\begin{eqnarray} \label{3.39}
\hat{a}^{\dag}_{\textbf{F}}\hat{a}_{\textbf{F}}=\hat{N}_{\textbf{F}}
\end{eqnarray}
where $\hat{N}_{\textbf{F}}$ represents the number operator for the
particles in the coherent state. Because of
$\langle\textbf{y}|\hat{N}_{\textbf{F}}|\textbf{y}\rangle=y^{*}_{\textbf{F}}y_{\textbf{F}}\neq0$
and the anti-commuting properties of Grassmann variables (Eq. 3.1),
it follows
\begin{eqnarray} \label{3.39}
\frac{\langle\textbf{y}|\{\hat{a}_{\textbf{F}},\hat{a}^{\dag}_{\textbf{F}}\}|\textbf{y}\rangle}{V}
=\frac{1+\{y_{\textbf{F}},y^{*}_{\textbf{F}}\}}{V}=\frac{1}{V}\longrightarrow0
\end{eqnarray}
This leads to the natural starting point of what we may call the
Grassmann-Bogoliubov approximation. Here one can identify
$y^{*}_{\textbf{F}}y_{\textbf{F}}=N_{\textbf{F}}(\approx \;N)$ as
the average particle number
of the state of macroscopic coherence in the thermodynamic limit.\\
\indent One point is to be noted here.
One may conclude that anti-commutation obeyed by fermions do not have
a classical analogy since they do not fulfill classical Poisson brackets as obeyed by
bosonic commutation relations. But this should not lead to misunderstanding.
The number operator
$\hat{N}=\sum_{k}\hat{N}_{k}$ and the Hamiltonian operator
$\hat{H}=\sum_{k}\epsilon_{k}\hat{N}_{k}$ have classical limits
because they are bilinear in $\hat{a}_{k},\;\hat{a}^{\dag}_{k}$.
For the particles obeying Fermi-Dirac statistics, only
quantities like charge, energy, current density or number density are measurable
classically because they are bilinear combination of field amplitudes
$\hat{a}_{k},\;\hat{a}^{\dag}_{k}$ \cite{18}. The amplitudes of the Fermi field is linear in
$\hat{a}_{k},\;\hat{a}^{\dag}_{k}$ and
can not be measured classically. Extrapolating the idea a bit
further we can emphasize 
that Grassmann fields themselves and fermionic field operators are,
by construction, fermionic while a product of even number of fermionic
quantities or Grassmann variables is bosonic which makes it
experimentally relevant \cite{15,18,30}.\\
\indent With this, we now return to our starting Hamiltonian (Eq.
2.5) characterizing the weakly interacting Fermi atoms. In the next
section the energy spectrum is calculated order by order under
Grassmann-Bogoliubov transformation.
\section{{Quasi-particle Spectrum and Quantum Fluctuations:}}
\subsection{{Lowest-order approximation:}}
\indent In the first approximation, we can neglect all the terms in
the Hamiltonian (Eq. 2.5) containing the operators $\hat{a}_{k}$ and
$\hat{a}^{\dag}_{k}$ with $k\neq \textbf{F}$. This implies that the
quantum fluctuation $(\delta\hat{\psi}(r))$ in Eq. (2.5) can be
ignored. Under such approximation the coherent state must be
occupied by macroscopically large number of particles i.e.
$N_{\textbf{F}}\sim N$. The replacement of $\hat{a}_{\textbf{F}}$
and $\hat{a}^{\dag}_{\textbf{F}}$ by $`y_{\textbf{F}}e^{i\theta}$' and
$`y^{*}_{\textbf{F}}e^{-i\theta}$' respectively, then becomes quite straight
forward. This substitution can not be made for a realistic potential
since it would result in a poor approximation at short distances of
order $r_{0}$, where the potential is strong and correlations are
important. The replacement is instead accurate in the case of a soft
potential whose perturbation is small at all distances \cite{27}.
The energy of the interacting system in the lowest order therefore
takes the form:
\begin{eqnarray} \label{3.39}
E_{\textbf{F}}=\frac{\hbar^{2}k^{2}_{\textbf{F}}}{2m}\hat{a}^{\dag}_{\textbf{F}}\hat{a}_{\textbf{F}}+
\frac{V_{0}}{2V}\hat{a}^{\dag}_{\textbf{F}}\hat{a}^{\dag}_{\textbf{F}}\hat{a}_{\textbf{F}}\hat{a}_{\textbf{F}}
=\frac{\hbar^{2}k^{2}_{\textbf{F}}}{2m}y^{*}_{\textbf{F}}y_{\textbf{F}}
+\frac{V_{0}}{2V}y^{*}_{\textbf{F}}y^{*}_{\textbf{F}}y_{\textbf{F}}y_{\textbf{F}}
\end{eqnarray}
To the same order of approximation, one can easily express the
parameter $V_{0}$ of Eq (4.1) in terms of the scattering
length $a$, using the result
$V_{0}=\frac{4\pi\hbar^{2}a}{m}$, under Born approximation.
However, due to the anti-commuting nature of the Grassmann variables
[Eq. (3.1)] the energy is given by only the first term in Eq. (4.1),
i.e., $E_{\textbf{F}}=
\frac{\hbar^{2}k^{2}_{\textbf{F}}}{2m}N_{\textbf{F}}=\mu N_{\textbf{F}}$, where,
$N_{\textbf{F}}=y^{*}_{\textbf{F}}y_{\textbf{F}}$ represents the
average number of particles and $`\mu$' is the chemical potential of the weakly interacting fremions in the coherent state.
\subsection{{Higher order approximation; calculation of quasi-particle spectrum:}}
\indent The result
$E_{\textbf{F}}=\mu N_{\textbf{F}}$
for the energy of the interacting fermions in the lowest order has
been obtained by taking into account of Eq (2.5) only for the
particle operators $\hat{a}_{k}$ and $\hat{a}^{\dag}_{k}$ with
$k=\textbf{F}$. The terms containing only one particle operator with
$k\neq \textbf{F}$ do not enter into the Hamiltonian (Eq. 2.5)
because of momentum conservation. By retaining all the quadratic
terms in the particle operators with $k\neq \textbf{F}$, in the next
higher order, we obtain the following decomposition of the
Hamiltonian;
\begin{eqnarray} \label{3.39}
\hat{\textbf{H}}=\sum_{k}\frac{\hbar^{2}k^{2}}{2m}\hat{a}^{\dag}_{k}\hat{a}_{k}+\frac{V_{0}}{2V}
\hat{a}^{\dag}_{\textbf{F}}\hat{a}^{\dag}_{\textbf{F}}\hat{a}_{\textbf{F}}\hat{a}_{\textbf{F}}
+\frac{V_{0}}{2V}\sum_{k\neq
\textbf{F}}\left(4\hat{a}^{\dag}_{\textbf{F}}\hat{a}^{\dag}_{k}\hat{a}_{\textbf{F}}\hat{a}_{k}+
\hat{a}^{\dag}_{k}\hat{a}^{\dag}_{-k}\hat{a}_{\textbf{F}}\hat{a}_{\textbf{F}}+
\hat{a}^{\dag}_{\textbf{F}}\hat{a}^{\dag}_{\textbf{F}}\hat{a}_{k}\hat{a}_{-k}\right)\hspace{0.25
cm}
\end{eqnarray}
\indent Now, the replacement of $\hat{a}^{\dag}_{\textbf{F}}$ and
$\hat{a}_{\textbf{F}}$ by $y^{*}_{\textbf{F}}e^{-i\theta}$ and $y_{\textbf{F}}e^{i\theta}$,
as carried out previously, yields the following expression for the
Hamiltonian;
\begin{eqnarray} \label{3.39}
\hat{\textbf{H}}=\sum_{k}\frac{\hbar^{2}k^{2}}{2m}\hat{a}^{\dag}_{k}\hat{a}_{k}+
\frac{V_{0}}{2V}y^{*}_{\textbf{F}}y^{*}_{\textbf{F}}y_{\textbf{F}}y_{\textbf{F}}
+\frac{V_{0}}{2V}\sum_{k\neq
\textbf{F}}\left(4y^{*}_{\textbf{F}}\hat{a}^{\dag}_{k}y_{\textbf{F}}\hat{a}_{k}+
\hat{a}^{\dag}_{k}\hat{a}^{\dag}_{-k}y_{\textbf{F}}y_{\textbf{F}}+
y^{*}_{\textbf{F}}y^{*}_{\textbf{F}}\hat{a}_{k}\hat{a}_{-k}\right)
\end{eqnarray}
The third term of this Hamiltonian represents the self-energy of the
excited states due to the interaction, simultaneous creation of the
excited states at momenta $k$ and $-k$ and simultaneous annihilation
of the excited states, respectively. But, the simultaneous creation
and annihilation of the excited states do not contribute to the
Hamiltonian ($\hat{\textbf{H}}$) due to the Grassmannian nature of
the fermionic field variables. Introducing the relevant interaction
coupling constant $g$ fixed by the s-wave scattering length $a$ as
\begin{eqnarray} \label{3.39}
g=\frac{4\pi\hbar^{2}a}{m}
\end{eqnarray}
the Hamiltonian (Eq. 4.3) can be written as
\begin{eqnarray} \label{3.39}
\hat{\textbf{H}}=E_{\textbf{F}}+\frac{1}{2}{\sum_{k}}'(E_{k}+2\rho
g)(\hat{a}^{\dag}_{k}\hat{a}_{k}+\hat{a}^{\dag}_{-k}\hat{a}_{-k})
\end{eqnarray}
Here ${\sum_{k}}'$ sign indicates that the terms $k=\textbf{F}$ are
omitted from the summation. $\rho$ represents the fermionic density
expressed as
$\rho=\frac{y^{*}_{\textbf{F}}y_{\textbf{F}}}{V}=\frac{N_{\textbf{F}}}{V}(\approx\frac{N}{V})$.
The physical coupling constant $g$ renormalizes the effective
potential $V_{0}$. $E_{k}$ refers to the energy when the
interaction $g=0$ and is given by $E_{k}=\frac{\hbar^{2}k^{2}}{2m}$. Unlike the bosonic case, the Hamiltonian in Eq (4.5) is peculiar in a sense that it has no terms with creation operators only or annihilation operators only. In other words, bosonic Hamiltonian does not conserve the number of particles. On the other hand in all physical contexts that have been explored experimentally, the number of fermions or more generally, the number of fermions minus the number of antifermions is strictly conserved \cite{15}. This conservation law leads to further restriction on the permissible states of the field. If a system starts with a state of fixed number of fermions, the conservation law restricts the set of accessible states considerably more than the $2\pi$ superselection rule mentioned earlier. Transitions can not be made, for example with different even number of fermions or between states with different odd number of fermions.\\   
\indent We now look for a solution of the problem (Eq. 4.5), i.e.,
energy eigenvalues of the Hamiltonian. Since the coherent state, represents a
combination of unperturbed eigenfunctions, neither $\hat{a}_{k}$ nor
$\hat{a}^{\dag}_{k}$ annihilate this state. The problem can be
solved exactly by a canonical transformation, namely
Grassmann-Bogoliubov transformation. To this end we introduce new
operators
\begin{eqnarray} \label{3.39}
\hat{a}_{k}&=&u_{k}\hat{A}_{k}+v^{*}_{-k}\hat{A}^{\dag}_{-k}\nonumber\\
\hat{a}^{\dag}_{k}&=&u^{*}_{k}\hat{A}^{\dag}_{k}+v_{-k}\hat{A}_{-k}
\end{eqnarray}
This transformations introduces a new set of operators $\hat{A}_{k}$
and $\hat{A}^{\dag}_{k}$ on which we impose the same fermionic
anti-commutation relation
\begin{eqnarray} \label{3.39}
\{\hat{A}_{k},\hat{A}^{\dag}_{k'}\}=\delta_{k,k'}
\end{eqnarray}
as obeyed by the original particle operators $\hat{a}_{k}$ and
$\hat{a}^{\dag}_{k}$, to make the transformation canonical.\\
\indent It can be easily verified that Eq. (4.7) are satisfied if
\begin{eqnarray} \label{3.39}
|u_{k}|^{2}+|v_{-k}|^{2}=1\;\;(\text{for any k})
\end{eqnarray}
Therefore $u_{k}$ and $v_{-k}$ can be chosen parametrically as
\begin{eqnarray} \label{3.39}
u_{k}=\cos\theta_{k}\;\;\;\;\;\text{and}\;\;\;\;\;\;v_{-k}=\sin\theta_{k}
\end{eqnarray}
By inserting Eq (4.6) into Eq. (4.5) we obtain
\begin{eqnarray} \label{3.39}
\hat{\textbf{H}}=E_{\textbf{F}}+\frac{1}{2}{\sum_{k}}'(E_{k}+2\rho
g)2|v_{-k}|^{2}+\frac{1}{2}{\sum_{k}}'(E_{k}+2\rho
g).(|u_{k}|^{2}-|v_{-k}|^{2}).(\hat{A}^{\dag}_{k}\hat{A}_{k}+\hat{A}^{\dag}_{-k}\hat{A}_{-k})
\nonumber\\+\frac{1}{2}{\sum_{k}}'(E_{k}+2\rho
g)2u_{k}v_{-k}(\hat{A}^{\dag}_{k}\hat{A}^{\dag}_{-k}+\hat{A}_{k}\hat{A}_{-k})\hspace{2
cm}
\end{eqnarray}
In order to make $\hat{\textbf{H}}$ diagonal in $\hat{A}_{k}$ and
$\hat{A}^{\dag}_{k}$ we use the freedom to eliminate the last term
of Eq. (4.10) i.e.,
\begin{eqnarray} \label{3.39}
\frac{1}{2}{\sum_{k}}'(E_{k}+2\rho g)2u_{k}v_{-k}=0
\end{eqnarray}
With Eq. (4.9) and by defining $\frac{1}{2}(E_{k}+2\rho g)$ as
$\alpha_{k}$, Eq (4.11) gives for each k-th mode
\begin{eqnarray} \label{3.39}
\alpha_{k}\sin2\theta_{k}=0
\end{eqnarray}
The above condition is satisfied both for positive to negative
values of $\alpha_{k}$ . The allowed values of the coefficient
$\theta_{k}$ are
\begin{eqnarray} \label{3.39}
\theta_{k}=\pm\frac{m\pi}{2}\;\;\;\;\;\;\;(m\text{=integer})
\end{eqnarray}
It is easy to note that for the above $\theta_{k}$ values, the first
term of Eq (4.10) corresponding to $\bar{E}$ we obtain
\begin{eqnarray} \label{3.39}
\bar{E}&=&E_{\textbf{F}}+\frac{1}{2}{\sum_{k}}'(E_{k}+2\rho
g)2|v_{-k}|^{2}\nonumber\\
&=&E_{\textbf{F}}+\frac{1}{2}{\sum_{k}}'(E_{k}+2\rho
g)2\sin^{2}2\theta_{k}
\end{eqnarray}
Since $\frac{1}{2}{\sum_{k}'}(E_{k}+2\rho g)2\sin^{2}2\theta_{k}=0$
we have $\bar{E}=E_{\textbf{F}}$. This is the same result in the
lowest order approximation obtained earlier. So the
$\hat{\textbf{H}}$ in Eq. (4.10) is modified as
\begin{eqnarray} \label{3.39}
\hat{\textbf{H}}=E_{\textbf{F}}+\frac{1}{2}{\sum_{k}}'(E_{k}+2\rho
g).(|u_{k}|^{2}-|v_{-k}|^{2}).(\hat{A}^{\dag}_{k}\hat{A}_{k}+\hat{A}^{\dag}_{-k}\hat{A}_{-k})
\end{eqnarray}
Identifying $\frac{1}{2}(E_{k}+2\rho
g).(|u_{k}|^{2}-|v_{-k}|^{2})=\alpha_{k}\cos2\theta_{k}=\alpha_{k}$,
the Hamiltonian in its final form can be written as
\begin{eqnarray} \label{3.39}
\hat{\textbf{H}}&=&E_{\textbf{F}}+\frac{1}{2}{\sum_{k}}'(E_{k}+2\rho
g)(\hat{A}^{\dag}_{k}\hat{A}_{k}+\hat{A}^{\dag}_{-k}\hat{A}_{-k})\nonumber\\
&=&E_{\textbf{F}}+\frac{1}{2}{\sum_{k}}'\boldsymbol\epsilon_{k}(\hat{A}^{\dag}_{k}\hat{A}_{k}+\hat{A}^{\dag}_{-k}\hat{A}_{-k})
\end{eqnarray}
where the quasi-particle energy $\boldsymbol\epsilon_{k}$ is
given by
\begin{eqnarray} \label{3.39}
\boldsymbol\epsilon_{k}=\frac{\hbar^{2}k^{2}}{2m}+2\rho
g=E_{k}+2\rho g
\end{eqnarray}
$E_{k}=\frac{\hbar^{2}k^{2}}{2m}$ stands for the free particle
energy if the interaction $g=0$. The operator
$\hat{A}^{\dag}_{k}\hat{A}_{k}$ resembles a particle number operator
and has eigenvalues $0$ and $1$. Hence the coherent state is
determined by the requirement that
\begin{eqnarray} \label{3.39}
\hat{A}_{k}|\textbf{y}(\theta)\rangle=0\;\;\;\;\text{for all}\;k\neq
\textbf{F}
\end{eqnarray}
Furthermore, all quasi-particle states correspond
to different numbers of non-interacting fermions. The expression
(4.16) is the fermionic counterpart of the Bogoliubov Hamiltonian
for bosons. $E_{\textbf{F}}$ term originating purely as kinetic
energy term for bosonic case is zero since the coherent state
corresponds to $k=0$ (instead of $k=\textbf{F}$). Secondly the
ground state is the vacuum of the quasiparticle operators
$\hat{A}_{k}$, $\hat{A}^{\dag}_{k}$. In general, the spectrum is
gapped \cite{14,30} everywhere corresponding to a non-zero
difference between the first excited state and the ground state. The
theory does not put any restriction upon the sign of $g$ and hence
it can be both positive and negative \cite{11,12,13,14}. For a critical negative $g$
i.e., $g_{c}$ it is possible that $\boldsymbol\epsilon_{k}$ becomes
zero when $E_{k}=-2\rho g_{c}$. In this case the spectrum can be
gapless \cite{30}. This is in consistent with the general perception that in the case of spontaneously broken continuous symmetry, there is always a low-lying excitation energy $\boldsymbol\epsilon_{k}$ which satisfies $\boldsymbol\epsilon_{k} \longrightarrow 0$ as its momentum $k \longrightarrow 0$. In the present context quasi-particle energy $\boldsymbol\epsilon_{k}$ smoothly goes to zero without a gap as the momentum goes to zero, even though the dispersion relation is quadratic in momentum rather than linear unlike the Bose gas.\\ 
\indent The coherent state $|\textbf{y}(\theta)\rangle$ is annihilated by
all $\hat{A}_{k},\;(k\neq \textbf{F})$. The transformation (4.6) can
be represented as
\begin{eqnarray} \label{3.39}
\hat{A}_{k}&=&\hat{\mathscr{U}}(\theta)\hat{a}_{k}\hat{\mathscr{U}}(\theta)^{\dag}\\
\hat{A}^{\dag}_{k}&=&\hat{\mathscr{U}}(\theta)\hat{a}^{\dag}_{k}\hat{\mathscr{U}}(\theta)^{\dag}
\end{eqnarray}
$\hat{\mathscr{U}}(\theta)$ represents the unitary displacement
operator which produces the
coherent state as given in Eq (3.13).
\begin{eqnarray} \label{3.39}
|\textbf{y}(\theta)\rangle=\hat{\mathscr{U}}(\theta)|0\rangle
\end{eqnarray}
The connection between $\hat{\mathscr{U}}(\theta)$ and
$\hat{{\textbf{D}}}(\textbf{y})$ in Eq. (3.13) can be established as
follows.
\begin{eqnarray} \label{3.39}
\hat{\mathscr{U}}(\theta)|0\rangle
&=&\hat{U}(\theta).\hat{{\textbf{D}}}(\textbf{y})|0\rangle\nonumber\\
&=&\hat{U}(\theta)|\textbf{y}\rangle\nonumber\\
&=&\exp\left(i\theta\hat{N}\right)|\textbf{y}\rangle\nonumber\\
&=&|e^{i\theta}\textbf{y}\rangle
\end{eqnarray}
Under the rotation of all Grassmann variables $y_{k}$ by
the same angle `$\theta$' i.e., the phase transformation of
$y_{k}\rightarrow e^{i\theta}y_{k}$ \cite{15} the displacement
operator produces altogether a different coherent state
$|e^{i\theta}\textbf{y}\rangle$. But for a physical rotation
$``\theta$'' is allowed to run from $0$ to $2\pi$. So, the
permissible values of $`\theta$' for which Eqs (4.19) and (4.20) are
valid are $\theta=0,\pm\frac{\pi}{2},\pm\pi,\pm\frac{3\pi}{2}$ as
obtained from Eq. (4.13). These values of $`\theta$' are the same as
obtained earlier by Braungardt $\textit{et. al.}$ in the
context of Fermi gas in optical lattice in 1d \cite{30}. For these specific
choices of $`\theta$', the phase of the coherent state gets stabilized.
\section{{Conclusion}}
\indent In this article we have tried to understand the coherent state of a 
weakly interacting ultracold Fermi gas as a close analogue of Bose condensate.
Notwithstanding statistical differences, the coherent state
of fermions bear a close kinship with that for bosons. Since the
fermionic operators anticommute their eigenvalues are anticommuting
numbers. These numbers or Grassmann variables play an important part
in the formulation of the present theoretical scheme. The fermionic
coherent state is not invariant under rotation $e^{i\theta \hat{N}}$
with number operator $\hat{N}$ for fermions, while the Hamiltonian
remains unitarily invariant. For \textit{preferred choices of $\theta$, the
coherent and the rotated coherent state are therefore
macroscopically distinct states because of spontaneous symmetry
breaking}. The description of the coherent state of weakly interacting fermions
and the rules of Grassmann algebra allow us to consider the
appropriate thermodynamic limit of the system. Bogoliubov
approximation can be implemented within the framework of this scheme
to realize this coherent state as a state of macroscopic coherence of fermions.
Unlike the bosonic case the lowest order
approximation on the Hamiltonian describing the weakly interacting
fermions yields an energy contribution due to kinetic energy of the
particles only. The quasi-particle spectrum arising from the higher
order interaction exhibits in general, the gaps \cite{14,30}. Our
analysis reveals that the close parallels between the phase space
quasi-probabilities of the boson and fermions as pointed out by
Cahill and Glauber \cite{15} have deep rooted consequences in the
physics of ultracold degenarate Fermi gas.

\begin{Large}
Acknowledgements:
\end{Large}
Thanks are due to the Council of Scientific and Industrial Research,
Government of India and also to the Department of Science and
Technology, Government of India under Grant-No SR/ S2/ LOP-13/ 2009
for partial financial support.

\end{document}